\documentclass[aps,prd,twocolumn,superscriptaddress,nofootinbib,longbibliography,10pt]{revtex4-2}
\usepackage{etoolbox}
\usepackage{dcolumn}
\usepackage{changepage}
\usepackage{amsmath,amssymb,amsfonts,mathtools,bm}
\usepackage{mathrsfs,bbold}
\usepackage{graphicx}
\usepackage[colorlinks=true,urlcolor=blue,citecolor=red,linkcolor=blue]{hyperref}
\usepackage{natbib}
\usepackage{wrapfig}
\usepackage{flushend}
\usepackage{BOONDOX-cal}

\begin{document}
\title{Gravity mediated entanglement of phonons in Bose-Einstein condensates}
\author{Soham Sen}
\email{sensohomhary@gmail.com}
\affiliation{Department of Astrophysics and High Energy Physics, S. N. Bose National Centre for Basic Sciences, JD Block, Sector-III, Salt Lake City, Kolkata-700 106, India}
\author{Sunandan Gangopadhyay}
\email{sunandan.gangopadhyay@gmail.com}
\affiliation{Department of Astrophysics and High Energy Physics, S. N. Bose National Centre for Basic Sciences, JD Block, Sector-III, Salt Lake City, Kolkata-700 106, India}
\author{Vlatko Vedral}
\email{vlatko.vedral@physics.ox.ac.uk}
\affiliation{Clarendon Laboratory, University of Oxford, Park Road, Oxford OX1 3PU, United Kingdom}
\begin{abstract}
\noindent The eigenstates of two test-masses (where each test-mass is placed inside of a harmonic trap) separated by a distance, can get entangled where gravity acts as the mediator of entanglement and it has been argued  in \href{https://doi.org/10.48550/arXiv.2511.07348}{arXiv:2511.07348 [quant-ph]} that this entanglement of masses cannot be generated without the underlying quantum nature of gravity. In this work, we consider two non-relativistic Bose-Einstein condensates (formed inside of harmonic trap potentials with identical trapping frequencies) separated by a distance. We take a linearized quantum gravity model and investigate the generation of entanglement while gravitons serve as the mediator of entanglement. The entanglement is generated between the phonon modes of the two condensates, and we observe that for very low separation distance, the entanglement generated is significantly higher than that observed for the quantum gravity induced entanglement of masses or QGEM protocol; however, the fall of entanglement is faster than the two-particle case for two separated Bose-Einstein condensates. We observe that when the number of particles in the condensate is increased, the degree of entanglement for a smaller separation distance becomes substantially higher compared to the case discussed in \href{https://doi.org/10.1103/PhysRevD.105.106028}{Phys. Rev. D 105 (2022) 106028}, which allows for a more robust experimental proposal using this quantum gravity induced entanglement of phonons or QGEP protocol.
\end{abstract}
\maketitle
\section{Introduction}
\noindent The advent of quantum mechanics and the general theory of relativity marked the birth of a new era in theoretical as well as experimental physics. The quantum world, being completely different from the classical realm, has expanded our perception of reality. In nature, three of the four fundamental forces, namely electromagnetic theory, weak nuclear force, and strong nuclear force, can be completely explained using a quantum field theoretic description. However, when one attempts to write down a full quantum field theory of gravity, it is close to an impossible task, with the primary problem being that the quantum field theory of gravity is non-renormalizable. The more straightforward way to look for the quantum nature of gravity is therefore to search for low-energy signatures of quantum gravity where we do not need to probe the Planck length scale. 

\noindent It is well-known that the mediator of gravity is a spin-two particle known as a graviton \cite{SNGupta1,SNGupta2,Feynman,DeWitt,DeWitt2}.
The perturbative quantum field theory of gravity is a very good effective field theory, as has been discussed in \cite{Donoghue}, irrespective of having subtle issues. To test the quantum nature of gravitons at low energies, a proposal of investing the spin entanglement between the two quantum superposed test masses has been given in \cite{Bose1,Bose2,Marletto}. In this quantum gravity induced entanglement of masses, one needs to create a spatial quantum superposition of two test masses and bring them close to each other. The experimental parameter needs to be controlled in this experiment in such a way that only the two masses interact via the exchange of massive gravitons. Later on, the entanglement generation due to gravitons has been investigated in detail in several literatures \cite{Graviton_1,Graviton_2,Graviton_3,
Graviton_4,Graviton_5,Graviton_6}. As has already been discussed in detail in \cite{Mechanism_Gravitons}, local operation and classical communication (LOCC) cannot generate entanglement \cite{LOCC}; however, it is possible via the use of local operation and quantum communication (LOQC). It was explicitly shown in \cite{Mechanism_Gravitons} that a classical gravitation field cannot generate entanglement between two test masses, and the entanglement generation between two masses placed inside harmonic trap potentials has been observed when the gravitational fluctuations are quantized. It was found that the energy shift in the gravitational field becomes an operator-valued interaction, and then finally the concurrence is calculated to obtain an analytical expression for the degree of entanglement generation between the matter systems. It was observed that the degree of entanglement increases with the increase in the masses for the matter systems and decreases with the increase in the distance between the two test masses and the frequency of the isotropic harmonic oscillators.

\noindent In our work, we have considered two Bose-Einstein condensates placed inside two harmonic traps separated by a finite distance. Instead of considering entanglement between the energy states of the condensate systems, we instead consider the entanglement generation between the quasi-particle excited states or the phonon states in this scenario. Even if excitations happen, the Bose-Einstein condensate does not get destroyed while working with quasi-particle excited states. Writing down the energy-momentum tensor for the model system, we then analytically obtain the energy shift in the gravitational field, which becomes operator valued for the consideration of graviton interaction with the Bose-Einstein condensates (BECs). We then perturbatively obtain the graviton-mediated entanglement generation for the model system and investigate whether Bose-Einstein condensates result in an increase in the degree of entanglement generation between the two BECs compared to the two-particle matter system considered in \cite{Mechanism_Gravitons}. 

\section{Phonon-phonon entanglement generation}
\noindent We start by considering a simple model of two systems with identical phonon mode frequencies $(\omega)$. Before going into the detailed quantum field theoretic construction of the BEC-graviton interaction model, we will go through an intuitive back of the envelop calculation to see whether phonon-phonon entanglement can actually be measurable in a realistic scenario. The energy of $n_1$ and $n_2$ number of phonons with same frequency for the first and second systems  can be expressed as
\begin{equation}\label{1.1}
E_{1,2}=\left(n_{1,2}+\frac{1}{2}\right)\hbar \omega.
\end{equation}
The mass then is expressed simply by $m_{1,2}=\frac{\hbar\omega}{c^2}\left(n_{1,2}+\frac{1}{2}\right)$. We consider simple interaction generated between the phonons because of there effective masses and the Newtonian potential for these two systems separated by a finite separation $d$ reads
\begin{equation}\label{1.2}
U=-\frac{G m_1 m_2}{d}=-\frac{G \hbar^2\omega^2}{d c^4}\left(n_1+\frac{1}{2}\right)\left(n_2+\frac{1}{2}\right)~.
\end{equation}
The total phase accumulation then is equal to $\phi=\frac{Ut}{\hbar}=-\frac{G \hbar\omega^2 t}{d c^4}\left(n_1+\frac{1}{2}\right)\left(n_2+\frac{1}{2}\right)$ and we obtain the effective time for obtaining maximal entanglement to be 
\begin{equation}\label{1.3}
\tau=\left\lvert\frac{\hbar}{U}\right\rvert=\frac{dc^4}{\hbar G \omega^2}\left(n_1+\frac{1}{2}\right)^{-1}\left(n_2+\frac{1}{2}\right)^{-1}~.
\end{equation}
The important thing to note is that $\tau$ is very large which can only be controlled provided $\omega$, $n_1$, $n_2$ to be large as well as $d$ to be very small. Consider a pair of systems where $n_1$ and $n_2$ are very large giving the effective time scale to be $\tau\simeq\frac{dc^4}{n_1 n_2\hbar G \omega^2}\sim 10^{78} \frac{d}{n_1n_2\omega^2}$ sec which allows for a significant reduction in the effective time scale. We can now check for the relative phase generated while we
consider the initial state of the system to be tensor product of two superposition state in the pair of systems as
$|\psi(0)\rangle=\frac{1}{\sqrt{2}}\left(|0\rangle_1+|1\rangle_1\right)\otimes \frac{1}{\sqrt{2}}\left(|0\rangle_2+|1\rangle_2\right)~.$ The simple interaction Hamiltonian structure from eq.(\ref{1.2}) can be constructed as
$\hat{U}=-\frac{G \hbar^2\omega^2}{d c^4}\left(\hat{n}_1+\frac{1}{2}\right)\left(\hat{n}_2+\frac{1}{2}\right)~.$
The time evolved state $|\psi(t)\rangle=e^{-\frac{i\hat{U}t}{\hbar}}|\psi(0)\rangle$ then takes the form 
\begin{equation}\label{1.4}
\begin{split}
|\psi(t)\rangle=&\frac{e^{i\Omega_0 t}}{2}\left[|0,0\rangle+e^{2i\Omega_0 t}[|0,1\rangle+|1,0\rangle]+e^{8i\Omega_0 t}|1,1\rangle\right]
\end{split}
\end{equation}
where $\Omega_0\equiv \frac{\hbar \omega^2 G}{4 d c^4}$. The relative phase is then $\phi_{11}-\phi_{00}=\frac{2\hbar \omega^2 G}{d c^4}$ which again is extremely small. The relative phase can be much higher if the superposition is considered between higher energy levels which may allow for a detection scenario. The important thing to notice is that considering standard phonon-phonon interaction in quantum mechanical systems may lead to serious difficulties in the maximal entanglement generation as well as detecting the relative phase in the final entangled state. The idea is therefore investigate entanglement generation between phonons where the systems allows for a large number of particle accumulation in a single energy level implying large numbers of single mode phonon accumulation as is evident from the expression of the time scale in eq.(\ref{1.3}). It is also important to use phonons with substantially high frequencies. For an experimental scenario scenario, we thus consider two Bose-Einstein condensates prepared in harmonic trap potential separated by a finite separation $d$. Before going into the details of the model we should first discuss the analytical expression for the order parameter in the presence of small amplitude oscillations.
\subsection{Gross-Pitaevski equation}\label{S2}
\noindent In this section, we shall start by considering the Gross-Pitaevski equation and the ground state solution of a weakly-interacting Bose-gas and then consider small amplitude oscillations with respect to the ground state of the Bose gas representing the Bose-Einstein condensate.

\noindent We start by considering the Hamiltonian for a single Bose-Einstein condensate placed inside of a harmonic trap potential given as
\begin{equation}\label{2.5}
H=\sum\limits_{k=1}^N\left(-\frac{\hbar^2}{2m}\frac{\partial^2}{\partial \vec{r}_k^2}+\frac{1}{2}m \omega^2 \vec{r}_k^2\right)+\sum\limits_{k<l}\frac{4\pi\hbar^2 \mathcal{a}}{m}\delta(\vec{r}_k-\vec{r}_l)
\end{equation}
where $N$ denotes the number of bosons, $\mathcal{a}$ denotes the scattering length, $m$ denotes the mass of each bosons, and the second term in the right hand side of the above equation denotes the two-particle interaction term. Making use of the variational principle one can obtain the single particle Gross-Pitaevski equation given by
\begin{equation}\label{2.6}
\left(-\frac{\hbar^2}{2m}\frac{\partial^2}{\partial \vec{r}^2}+\frac{1}{2}m \omega^2 \vec{r}^2+\frac{4\pi\hbar^2 \mathcal{a}}{m}|\psi(\vec{r})|^2\right)\psi(\vec{r})=\mu \psi(\vec{r})
\end{equation}
where $\psi(\vec{r})$ denotes the ground state wave function for the system that minimizes the expectation value of the Hamiltonian in eq.(\ref{2.5}), and $\mu$ denotes the corresponding chemical potential for the system. As has been shown in \cite{BEC_Pitaevski_Stingari} and also used in \cite{BEC_PRX}, one can obtain a simplified solution of the ground state wave function by setting $g=0$ instead of using $g\equiv\frac{4\pi\hbar^2 \mathcal{a}}{m}$ and the chemical potential is set to $\mu=\frac{1}{2}\hbar\omega$. The ground state wave function after normalization then reads (for bosons inside of a non-interacting Bose-Einstein condensate) \cite{BEC_Pitaevski_Stingari}
$\psi_0(\vec{r})=\left(\frac{m\omega}{\pi \hbar}\right)^{\frac{3}{4}}\exp\left[-\frac{m\omega r^2}{2\hbar}\right].$ We are now in a position to investigate the analytical expression for the quantized order parameter of the Bose-Einstein condensate.
\subsection{The order parameter and small amplitude oscillations}
\noindent For a bosonic system, the field operator $\hat{\psi}(r)$ can be expressed as
\begin{equation}\label{2.7}
\hat{\psi}(\vec{r})=\sum_k\hat{a}_k\psi_k=\hat{a}_0\psi_0+\sum_{k\neq 0}\hat{a}_k\psi_k
\end{equation}
where $\hat{a}_k$ annihilates the vacuum state corresponding to the $k$-th mode. For a Bose-Einstein condensate all the particles take the ground state of the system and the ground state portrays an overall macroscopic behaviour as a result of the matter wave overlap. Now, $\langle \hat{a}_0^\dagger\hat{a}_0\rangle=N_0$ with $N_0$ denoting the number of particles in the ground state of the system ($N_0\gg 1$) where one can implement the Bogoliubov approximation by considering $\hat{a}_0\sim\sqrt{N_0}$ which enables one to write down the order parameter as
\begin{equation}\label{2.8}
\hat{\psi}(\vec{r})\simeq\sqrt{N}_0\psi_0(\vec{r})+\delta \hat{\psi}(\vec{r})
\end{equation}
with the definition of the excitation term as $\delta \hat{\psi}(\vec{r})=\sum_{k\neq 0}\hat{a}_k\psi_k$. Here, we shall be primarily interested in small amplitude oscillations where $\delta\hat{\psi}(\vec{r})$ denotes the small perturbations with respect to the ground state of the Bose gas denoting the condensate state. This small amplitude oscillations can sometimes be denoted by collective excitations behaving as a particle which are also known as quasiparticles and the corresponding states are known as quasiparticle states. This collective small amplitude fluctuations, in our model, describe Bogoliubov-quasiparticles or phonons. For a fixed mode $k$, we can express the quasi-particle vacuum $|0\rangle_k$ as $\hat{b}_k|0\rangle_k=0$
where the quasiparticle raising and lowering operators satisfy the following commutation relation as
\begin{equation}\label{2.9}
\begin{split}
[\hat{b}_k,\hat{b}^\dagger_{k'}]=\delta_{k,k'}~,~~[\hat{b}_k,\hat{b}_{k'}]=[\hat{b}_k^\dagger,\hat{b}_{k'}^\dagger]=0~.
\end{split}
\end{equation}
Considering positive energy modes only, we can finally write down the position dependent part of the order parameter after a Bogoliubov transformation to be
\begin{equation}\label{2.10}
\hat{\Psi}(x)=\sqrt{N_0}\left(\frac{m\omega}{\pi\hbar}\right)^\frac{1}{4}e^{-\frac{m\omega x^2}{2\hbar}}\left(1+\varepsilon\hat{b}_k\right)
\end{equation}
with the dimensionless $\varepsilon$ being defined as $\varepsilon\equiv\frac{1}{\sqrt{N_0}}$ which indeed is a very small quantity as $N_0\gg 1$. As we shall be considering only single quasi-particle or phonon mode oscillations, it is quite straightforward to set $\hat{b}_k=\hat{b}$. In the next section, we shall explain in details the model considered in this analysis\footnote{For a detailed derivation of eq.(\ref{2.10}) see the supplementary material}.
\section{Gravity mediated entanglement generation}\label{S3}
\noindent With the motivation and initial understanding in place, we can now proceed towards considering the set-up in our model system and from their write down the interaction Hamiltonian to calculate the gravity mediated entanglement generation.
\subsection{Two BECs and the background model}
\begin{figure}
\begin{center}
\includegraphics[scale=0.9]{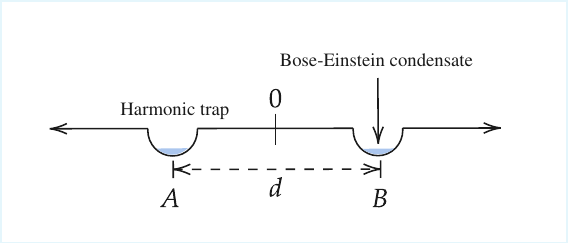}
\caption{Two Bose-Einstein condensates are generated inside of two harmonic traps such that they are separated by a finite spatial distance $d$.\label{BEC_Joined_State}}
\end{center}
\end{figure}
\noindent We consider two harmonic traps at points $A$ and $B$ where their centre of masses are separated by a distance $d$. In these two traps with identical trap frequencies $\omega$, two Bose-Einstein condensates are created. 
We consider the ground stater of the combined system given in Fig.(\ref{BEC_Joined_State}) as a tensor product state of two Bogoliubov vacuum states corresponding to the BECs created at $A$ and $B$. The joint Bogoliubov vacuum state can then be expressed as
\begin{equation}\label{2.11}
|\psi_0\rangle=|0\rangle_A\otimes|0\rangle_B~.
\end{equation}
This Bogoliubov vacuum states simply represent the quasiparticle vacuum state or a BEC with $N_0$ number of particles with no quasiparticles in it \cite{Bogoliubov_Vacuum,Bogoliubov_Vacuum_2}. Before progressing further, we need to start considering the model Hamiltonian for the system.  The matter part of the Hamiltonian can then be expressed as
\begin{equation}\label{2.12}
\hat{H}_{\text{Mat}}=\hat{H}_A\otimes \hat{\mathbb{1}}_A+\hat{\mathbb{1}}_B\otimes\hat{H}_B+\lambda\hat{H}_{AB}
\end{equation}
where the coupling $\lambda$ is small. It is therefore always prudent to apply perturbation theory in order to obtain the final state of the system.  At first, one needs to calculate the correction to the energy of the ground state for the coupling term $\lambda \hat{H}_{AB}$, and then one shall calculate the correction to the ground state of the system. If the ground state of the system reads $|\psi_0\rangle$ then we assume that the perturbed ground state of the system can be expressed as $|\psi_0^\lambda\rangle=|\psi_0\rangle+\lambda|\psi^{(1)}\rangle$ with $|\psi^{(1)}\rangle$ denoting the first order correction to the ground state.
Up to order $\mathcal{O}(\lambda^0)$, it is possible to obtain the relation $E^{(0)}=E_A^{(0)}+E_B^{(0)}$ where $\hat{H}_A|0\rangle_A=E_A^{(0)}|0\rangle_A$ and $\hat{H}_B^{(0)}|0\rangle_B=E_B^{(0)}|0\rangle_B$. The first order correction to the energy is obtained from the equation $\hat{H}_{\text{Mat}}|\psi^\lambda_0\rangle=E^\lambda|\psi^\lambda_0\rangle$ when we compare the components up to $\mathcal{O}(\lambda)$. Here, $E^\lambda=E^{(0)}+\lambda E^{(1)}$ and the first order correction to the ground state energy is obtained as $E^{(1)}=\langle \psi_0|\hat{H}_{AB}|\psi_0\rangle $. Following this time-independent perturbation theory, one can also obtain the first order correction to the ground state of the system as $|\psi^{(1)}\rangle=\sum_{n_A,n_B}C_{n_An_B}|n_A,n_B\rangle$ where $n\neq0$. 
As we are considering identical Bose-Einstein condensates at positions $A$ and $B$, $E_A^{(0)}=E_B^{(0)}=E_0$ which helps us to define the connection coefficients as
\begin{equation}\label{2.13}
C_{n_An_B}\equiv\frac{\langle n_A,n_B|\hat{H}_{AB}|0_A,0_B\rangle}{2E_0-E_{n_A}^{(0)}-E_{n_B}^{(0)}}~.
\end{equation}
The total state of the model system then reads
$|\psi_0^{\lambda}\rangle=\frac{1}{\mathcal{N}}\left(|\psi_0\rangle+\lambda|\psi^{(1)}\rangle\right)$
with $\mathcal{N}$ being the normalization constant satisfying
$\mathcal{N}^2=1+\lambda^2\sum_{n_A,n_B}|C_{n_A,n_B}|^2~.$
As also has been depicted in \cite{Mechanism_Gravitons}, in our case also the leading order entanglement generation occurs at $\mathcal{O}(\lambda)$. To investigate entanglement generation in the quantum state, we first need the interaction Hamiltonian for the system.

\subsection{Interaction Hamiltonian for the model system}
\noindent In this subsection, we proceed to obtain the interaction part of the Hamiltonian operator up to first order in the gravitational fluctuation which has the analytical form given by
\begin{equation}\label{2.14}
\hat{H}_{\text{int}}=-\frac{1}{2}\int d^3\vec{x}~ \hat{h}^{\mu\nu}(\vec{x})\hat{T}_{\mu\nu}(\vec{x})
\end{equation}
Following the analysis in \cite{SNGupta1,SNGupta2}, one can decompose $\hat{h}_{\mu\nu}$ around a Minkowski background as $\hat{h}_{\mu\nu}=\hat{\mathcal{h}}_{\mu\nu}-\frac{1}{2}\eta_{\mu\nu}\hat{\mathcal{h}}$
where $\mathcal{h}_{\mu\nu}$ denotes the spin-two modes whereas $\mathcal{h}=\eta^{\mu\nu}\mathcal{h}_{\mu\nu}$ denotes the spin-zero modes and can be treated as independent to each other \cite{Mechanism_Gravitons}. We shall primarily be focussing on the $\{0,0\}$ component of the energy-momentum tensor only. The interaction Hamiltonian in the scenario where the $\{0,0\}$ component of the energy momentum tensor dominates (the static limit) reads
\begin{equation}\label{2.15}
\begin{split}
\hat{H}_{\text{int}}&= \frac{1}{2}\int d^3\vec{x}\left(\hat{\mathcal{h}}_{00}(\vec{x})-\frac{1}{2}\eta_{00}\hat{\mathcal{h}}(\vec{x})\right)\hat{T}_{00}(\vec{x})~.
\end{split}
\end{equation}
One can consider $\hat{\mathcal{h}}_{\mu\nu}$ and $\hat{\mathcal{h}}$ as independent self-adjoint operators and execute a mode decomposition as \cite{Mechanism_Gravitons}
\begin{equation}\label{2.16}
\begin{split}
\hat{\mathcal{h}}_{\mu\nu}=&\kappa\int d^3\vec{k}\sqrt{\frac{\hbar}{2\omega_{\vec{k}}(2\pi)^3}}\left[\hat{\mathcal{H}}_{\mu\nu}(\vec{k})e^{-i\vec{k}\cdot\vec{x}}+\hat{\mathcal{H}}^\dagger_{\mu\nu}(\vec{k})e^{i\vec{k}\cdot\vec{x}}\right]
\end{split}
\end{equation}
where $\kappa\equiv \sqrt{\frac{16\pi G}{c^2}}$. Similarly the decomposition of the spin-zero operator takes the form
\begin{equation}\label{2.17}
\hat{\mathcal{h}}=2\kappa\int d^3\vec{k}\sqrt{\frac{\hbar}{2\omega_{\vec{k}}(2\pi)^3}}\left[\hat{\mathcal{H}}(\vec{k})e^{-i\vec{k}\cdot\vec{x}}+\hat{\mathcal{H}}^\dagger(\vec{k})e^{i\vec{k}\cdot\vec{x}}\right]~.
\end{equation}
The one particle state can be expressed as $|\vec{k}\rangle=\left(\hat{\mathcal{H}}^{\dagger}_{00}(\vec{k})+\hat{\mathcal{H}}^{\dagger}(\vec{k})\right)|0\rangle$. It is then easy to show that the first order correction to the energy $E_{\vec{k}}^{(1)}=\langle 0|\hat{H}_{\text{int}}|0\rangle=0$. Hence, graviton induced shift in the energy level can be observe at the second order only. The second order correction to the energy is obtained as
\begin{equation}\label{2.18}
\begin{split}
E_{\vec{k}}^{(2)}=&\langle \Psi_0|\hat{H}_{\text{int}}|\Psi^{(1)}\rangle
=-\sum_{\vec{k}\neq 0}\frac{\langle 0|\hat{H}_{\text{int}}| \vec{k}\rangle\langle \vec{k}|\hat{H}_{\text{int}}|0\rangle}{\hbar\omega_{\vec{k}}}~.
\end{split}
\end{equation}
where energy of the graviton state $|\vec{k}\rangle$ is $E_{\vec{k}}=E_0+\hbar\omega_{\vec{k}}$ and $(\hat{\mathcal{H}}_{00}(\vec{k})+\hat{\mathcal{H}}(\vec{k}))|0\rangle=|k\rangle$. Considering the continuum picture, it is now possible to write down the expression as
\begin{equation}\label{2.19}
\Delta \hat{H}_g=-\int d^3\vec{k}\frac{\langle 0|\hat{H}_{\text{int}}| \vec{k}\rangle\langle \vec{k}|\hat{H}_{\text{int}}|0\rangle}{\hbar\omega_{\vec{k}}}~.
\end{equation}
Using the analytical expression for the interaction Hamiltonian in eq.(\ref{2.15}) and the spin-2 and spin-0 graviton operators from eq.(s)(\ref{2.16},\ref{2.17}), we arrive at the expression for the shift in the energy of the graviton vacuum as (perturbation theory is executed up to  second order in the interaction Hamiltonian)
\begin{equation}\label{2.20}
\begin{split}
\langle 0|\hat{H}_{\text{int}}|\vec{k}\rangle&=\frac{\kappa}{2}\sqrt{\frac{\hbar}{2\omega_{\vec{k}}}}\hat{T}_{00}(\vec{k})~.
\end{split}
\end{equation}
\noindent To obtain the analytical expression for the above expression, we need to write down the analytical expression for the energy momentum tensor and obtain its Fourier transform. The dimension of the energy momentum tensor is $[T_{00}(\vec{r})]=ML^{-1}T^{-2}$ and that of the wave function $[\Psi(\vec{r})]=L^{-\frac{3}{2}}$ ($\int d^{3}\vec{r}~\Psi^{\dagger}(\vec{r})\Psi(\vec{r})=N_0$) where the analytical expression for the three-dimensional wave function operator of a Bose-Einstein condensate can be expressed as 
\begin{equation}\label{2.21}
\hat{\Psi}(\vec{r})=\sqrt{N}_0\left(\frac{m\omega}{\pi\hbar}\right)^{\frac{3}{4}}e^{-\frac{m\omega r^2}{2\hbar}}\left(1+\varepsilon \hat{b}_{\vec{k}}\right)
\end{equation}
with $\varepsilon\equiv \frac{1}{\sqrt{N_0}}\ll1$ as $N_0\gg 1$.
One can then write down the energy-momentum tensor for the model discussed in Fig.(\ref{BEC_Joined_State}) using the above expression for the wave function operator as
\begin{equation}\label{2.22}
\begin{split}
\hat{T}_{00}(\vec{r})= mc^2 \hat{\Psi}^{\dagger}(\vec{r})\hat{\Psi}(\vec{r})L^3\left(\delta(\vec{r}-\hat{\vec{r}}_A)+\delta(\vec{r}-\hat{\vec{r}}_B)\right)
\end{split}
\end{equation}
with $L$ being a length scale which fixes the dimension of the energy momentum tensor or the quantization length. Here, $\hat{\vec{r}}_A=\{\hat{x}_A,0,0\}$ and $\hat{\vec{r}}_B=\{\hat{x}_B,0,0\}$ where the operators $\hat{x}_A$ and $\hat{x}_B$ are given by $\hat{x}_A=\frac{d}{2}+\delta \hat{x}_A~\text{and}~\hat{x}_B=-\frac{d}{2}+\delta\hat{x}_B.$
The fluctuation operators $\delta \hat{x}_A$ and $\delta\hat{x}_B$ can be represented in terms of the quasiparticle (phonon) raising and lowering operators at the positions `A' and `B' respectively. For notational simplicity, we express the lowering operator at position `A' by $\hat{\alpha}$ and the lowering operator at position `B' by $\hat{\beta}$. In terms of these quasiparticle raising and lowering operators, we can express the fluctuations $\delta\hat{x}_A$ and $\delta\hat{x}_B$ by
\begin{equation}\label{2.23}
\delta\hat{x}_A=\sqrt{\frac{\hbar}{2m\omega}}(\hat{\alpha}+\hat{\alpha}^{\dagger})~\text{and}~\delta\hat{x}_B=\sqrt{\frac{\hbar}{2m\omega}}(\hat{\beta}+\hat{\beta}^{\dagger})~.
\end{equation} 
To proceed further, we need to take the Fourier transform of the energy-momentum tensor as
$\hat{T}_{00}(\vec{k})=\frac{1}{2\pi}\int d^3\vec{r}~\hat{T}_{00}(\vec{r})e^{-i\vec{k}\cdot\vec{r}}~.$
After quite a bit of analytical steps, we arrive at the lowest order matter-matter interaction term of the Hamiltonian as\footnote{For a detailed derivation refer to the supplementary material.} 
\begin{equation}\label{2.24}
\begin{split}
\Delta\hat{\mathcal{H}}_g^{AB}\simeq&\frac{\mathcal{A}_{N_0}^2\kappa^2}{16\pi c^2}\frac{e^{-\frac{m\omega d^2}{2\hbar}}}{d}\left[\frac{m\omega d^2}{2\hbar}+\frac{\hbar}{m\omega d^2}-\frac{5\varepsilon^2}{4}\right]\hat{\alpha}^\dagger\hat{\beta}^\dagger~.
\end{split}
\end{equation}
with $\mathcal{A}_{N_0}$ being defined as $\mathcal{A}_{N_0}\equiv N_0mc^2\left(\frac{m\omega L^2}{\pi\hbar}\right)^{\frac{3}{2}}$. 
We observe from the form of the interaction Hamiltonian that the standard $\frac{1}{d^3}$ contribution observed in \cite{Mechanism_Gravitons} has now been replaced by the leading order $d e^{-\frac{m\omega d^2}{2\hbar}}$ term. This leading term behaviour indeed signifies that the contribution will fall off pretty quickly if the two condensates are vastly separated and it will be impossible to obtain any signatures of graviton mediated entanglement.
\subsubsection{The entanglement time scale and phase shift}
\noindent Now, if we look at the interaction Hamiltonian in eq.(\ref{2.24}) of our model, we can read off the true potential to be
$H_{AB}\simeq \frac{\mathcal{A}_{N_0}^2\kappa^2}{16\pi c^2}e^{-\frac{m\omega d^2}{2\hbar}}\frac{m\omega d}{2\hbar}$. Then one can obtain the total phase accumulation in time $t$ to be $\phi_{AB}=\frac{H_{AB}t}{\hbar}=\frac{N_0^2 m^2 G t}{\hbar}\left(\frac{m\omega L^2}{\pi\hbar}\right)^3e^{-\frac{m\omega d^2}{2\hbar}}\frac{m\omega d}{\hbar}.$
The effective time for maximal entanglement generation is then
\begin{equation}\label{2.25}
\tau=\left(\frac{\pi\hbar}{m\omega L^2}\right)^3\frac{\hbar}{N_0^2 m^2 G}\frac{2\hbar}{m\omega d}e^{\frac{m\omega d^2}{2\hbar}}~.
\end{equation}
It is evident that for large separation between the harmonic traps $\tau\gg 1$, however, one can set $d\gtrsim\sqrt{\frac{2\hbar}{m\omega}}$ without violating the basic set-up of the model. For instance, if we set $d=2\sqrt{\frac{2\hbar}{m\omega}}$ then $\tau$ simply becomes
$\tau=\left(\frac{\pi\hbar}{m\omega L^2}\right)^3\frac{\hbar e^{4}}{2N_0^2 m^2 G}\sqrt{\frac{2\hbar}{m\omega}}~.$
As can be followed from the derivation, $m$ gives the mass of a single boson $m\sim 10^{-25}$ kg and $\omega$ can be set to $10^4$ Hz as a reference. If one increases the quantization length, it is possible to effectively reduce $\tau$ but for the worst case scenario, we consider here $L=\sqrt{\frac{\hbar}{m\omega}}$. Now, $N_0$ denotes the number of atoms in the condensate ($N_{0_1}=N_{0_2}$) which is quite large and can be as high as $10^{20}$. In a laboratory based set-up one can easily get $N_0\sim 10^9$ which gives $\tau$ to be equal to $\tau\sim 2\times 10^4$ sec. With $N_0\sim 10^{11}$ it is possible to obtain the effective scale of entanglement generation to be of the order of 2 sec but creating a condensate with $N_0\sim 10^{11}$ will be an experimental challenge in itself. One can also observe the relative phase shift to be simply of the order of $\Omega_{\text{Shift}}=\phi_{11}-\phi_{00}\sim\frac{N_0^2 m^2 G}{\hbar}\left(\frac{m\omega L^2}{\pi\hbar}\right)^3e^{-\frac{m\omega d^2}{2\hbar}}\frac{m\omega d}{\hbar}$ which is for the parameter values used earlier and $N_0\sim 10^9$ has an estimated value of $\Omega_\text{Shift}\simeq 5\times 10^{-4}$ $\text{sec}^{-1}$. This relative phase shift also is measurable in an experimental scenario. We shall now investigate the measure of entanglement for the QGEP protocol via calculating the concurrence and compare it with the QGEM result.
\subsection{Concurrence and the point-particle limit}
 \noindent We shall now calculate the concurrence for the model system. The state of the model system can be expressed as
\begin{equation}\label{2.26}
|\psi_f\rangle=\frac{1}{\mathcal{N}}\left(C_{0_A0_B}|0_A,0_B\rangle+C_{1_A1_B}|1_A,1_B\rangle\right)
\end{equation}
where $C_{0_A0_B}=1$ and $\mathcal{N}$ is the normalization constant. It is important to note that all components involving $|0_A,1_B\rangle$ and $|1_A,0_B\rangle$ states have been neglected. The normalization constant is obtained simply to be $\mathcal{N}=\sqrt{1+|C_{1_A1_B}|^2}$. The density matrix for the system is obtained by simply writing $\hat{\rho}_{AB}=|\psi_f\rangle\langle\psi_f|$. The reduced density matrix for the system is obtained by tracing out the `B' part of the quantum state as
\begin{equation}\label{2.27}
\begin{split}
\hat{\rho}_A=&\sum_{n_B}\langle n_B|\psi_f\rangle\langle\psi_f| n_B\rangle\\
=&\frac{1}{\mathcal{N}^2}\sum_{n_A,\bar{n}_A,n_B}C_{n_An_B}C^*_{\bar{n}_An_B}|n_A\rangle\langle\bar{n}_A|
\end{split}
\end{equation}
Using the above expression for the reduced density matrix, one can write down the analytical expression for concurrence as \cite{Concurrence1,Concurrence2} $\mathcal{C}\equiv\sqrt{2\left(1-\text{tr}\left[\hat{\rho}_A^2\right]\right)}.$
Using the analytical form of the final state from eq.(\ref{2.26}), we can obtain the expression for the concurrence to be
\begin{equation}\label{2.28}
\begin{split}
\mathcal{C}=&\frac{2|C_{1_A1_B}|}{1+|C_{1_A1_B}|}~.
\end{split}
\end{equation}
As $|C_{1_A1_B}|<1$, it is always possible to express the concurrence in a much simpler form as $\mathcal{C}\simeq2|C_{1_A1_B}|$ which has the analytical form given by
\begin{equation}\label{2.29}
\begin{split}
\mathcal{C}\simeq \frac{N_0^2m^2 G}{\hbar \omega d}\left[\frac{m\omega L^2}{\pi\hbar}\right]^3e^{-\frac{m\omega d^2}{2\hbar}}\left[\frac{m\omega d^2}{2\hbar}-\frac{5}{4N_0}+\frac{\hbar}{m\omega d^2}\right]
\end{split}
\end{equation}
where we have substituted $\mathcal{A}_{N_0}$.
The concurrence obtained in our analysis is vastly different from the model observed in \cite{Mechanism_Gravitons}. We shall now conduct a side by side comparison of the two models. 
\begin{figure}
\includegraphics[scale=0.28]{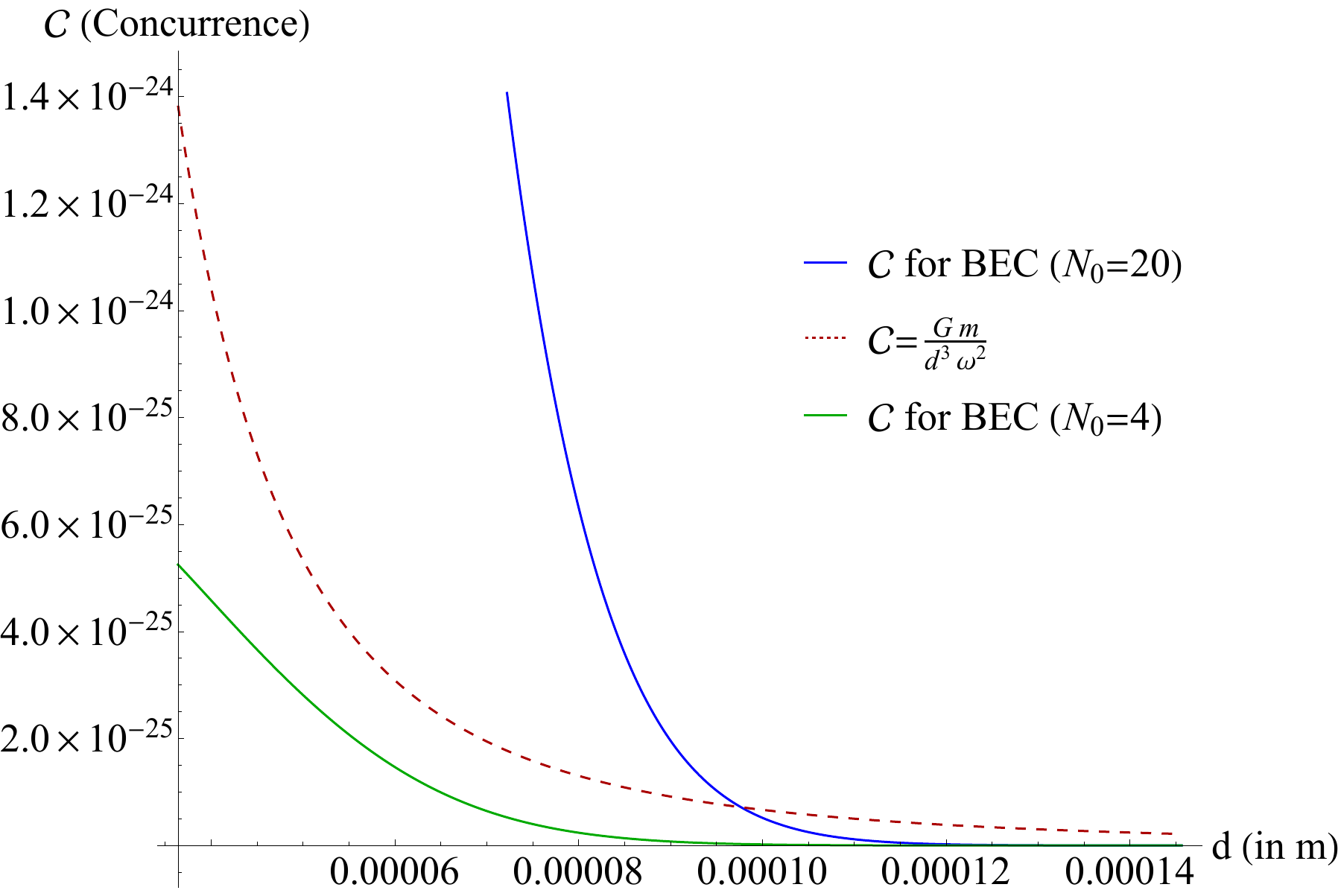}
\caption{Plot of Concurrence for the BEC model with different values of the boson numbers against the model in \cite{Mechanism_Gravitons} where $\mathcal{C}=\frac{G m}{d^3\omega^2}$.\label{Concurrence_Plot}}
\end{figure}
In order to plot this function against the distance between the centre of masses of two condensates, we fix the following values of the parameters, $m=10^{-25}$ kg and $\omega=10$ Hz. In Fig.(\ref{Concurrence_Plot}), we plot the concurrence of the model BEC system for different values of the boson numbers $N_0$ in the condensate against the distance between the centre of masses of the two condensates. We keep the quantization volume to be of the order of $L\sim \sqrt{\frac{\hbar}{m\omega}}$. We observe that when the number of bosons in the condensate system is very low ($N_0=4$), the concurrence is relatively smaller for $d\sim \sqrt{\frac{2\hbar}{m\omega}}$ or higher and it decays rapidly if the distance between the two condensates are kept on increasing. However, for a relatively large number of particles $(N-0=20)$, we observe that close to $d\sim \sqrt{\frac{2\hbar}{m\omega}}$, the concurrence is substantially larger than the one observed in \cite{Mechanism_Gravitons} for particles with mass $m\sim10^{-25}$ kg and again it decays rapidly with increase in $d$. This behaviour stays intact even with the increase in the frequency $\omega$ for $d\gtrsim \sqrt{\frac{2\hbar}{m\omega}}$. 
\begin{figure}[ht!]
\includegraphics[scale=0.28]{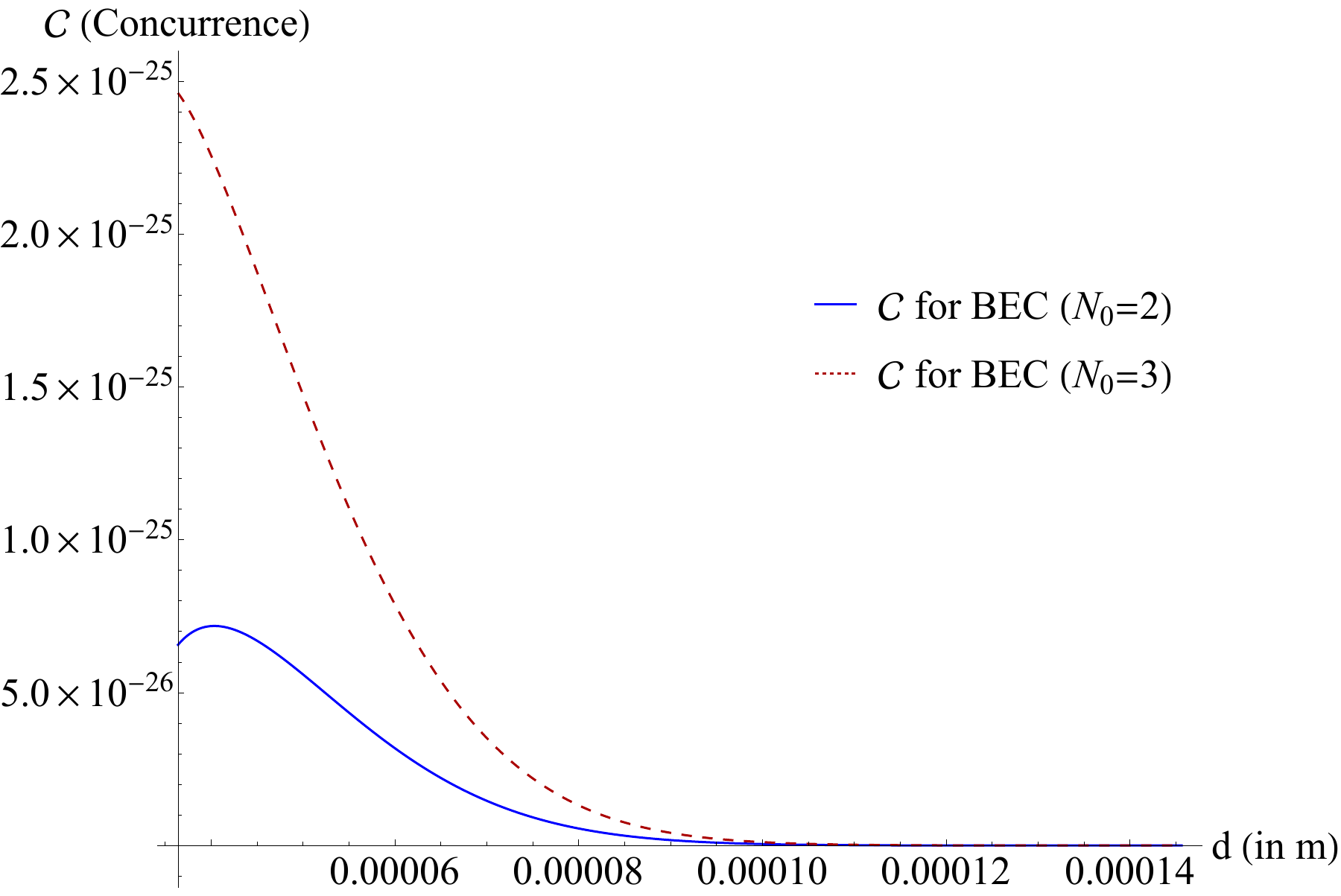}
\caption{Plot of Concurrence for the BEC model with very low values of the boson numbers $N_0$.\label{Concurrence_Plot_2}}
\end{figure}
In Fig.(\ref{Concurrence_Plot_2}), we again plot the concurrence for our BEC model for very small values of the Boson number $N_0$. As $N_0>1$, the minimum value it can attain is $N_0=2$ for which $\varepsilon<1$ gets satisfied. We observe that the degree of entanglement at first rises then decays with the increasing value of $d$. This initial rise in entanglement degradation from a lower value is a pure effect of the sub-leading $\frac{5}{4N_0}$ with a negative sign. Now, the quantization length $L$ has been considered to be of the order of $\sqrt{\frac{\hbar}{m\omega}}$ which is smaller than the minimum separation between the points `A' and `B'. This is quite valid as the length $L$ corresponds to individual condensates. For realistic models, $N_0\sim 10^9$ the concurrence value will be substantially higher leading to a more realistic detection scenario for this quantum gravity induced entanglement of phonons or QGEP protocol. We can also recover the result in \cite{Mechanism_Gravitons} by considering $d<\sqrt{\frac{\hbar}{m\omega}}$ ($=\sigma$ being the Gaussian width) and the quantization length $L\sim \sqrt{\frac{\pi\hbar}{m\omega}}$ which allows us to obtain the analytical result of the concurrence to be $\mathcal{C}\simeq \frac{N_0^2m^2G}{\hbar\omega d}(1-\frac{m\omega d^2}{2\hbar})(\frac{m\omega d ^2}{2\hbar}+\frac{\hbar}{m\omega d^2})\simeq \frac{N_0^2G m}{d^3\omega^2}$ and hence the $\mathcal{C}\propto \frac{G m}{d^3\omega^2}$ limit for smaller values of the separation term $d$ (such that $d$ is smaller than the Gaussian width of the Bose-Einstein condensate) and lower values of the total number of particles. The actual limit can be understood very clearly by actually taking the Gaussian width to be very large which is ensured by making $\omega$ to be very small. This scenario actually indicates complete delocalization of the wave function for obtaining the Newtonian quantum gravity induced potential in QGEM scenario.

\subsection{Experimental feasibility}
\noindent In our model of Bose-Einstein condensates separated by a finite distance, we observe that the degree of entanglement can be substantially enhanced by creating a Bose-Einstein condensate with higher and higher number of particles. Hence, creating an experimental model with two Bose-Einstein condensates will substantially increase the chances of detecting graviton mediated entanglement in massive systems. It is although important to note that the entanglement is not generated strictly between the centre of masses rather the entanglement is generated between the phonon modes. Now the important behaviour is that the high level of entanglement decays way faster due to the exponential decay factor and as a result the experimental implementation may be a bit challenging but at the same time more robust. The experimental model can be created by two condensates where one of the condensates can be taken away or nearer to the other condensates and if it is found out that the entanglement falls of rapidly while at low separation it rises very fast then it will be an evident evidence of the existence of graviton mediated entanglement. However, it is important to remember that the entanglement generation is primarily between the quasiparticle excited states. The exponential term actually considers the spatial overlap of the wave functions instead of considering only quantum superposition of the wave packets. This is a very strong and novel insight of the analysis presented in this manuscript.
\section{On the deflection of two parallel atom laser beams}
\noindent Till now, we have mostly concerned about the static part of the energy momentum tensor for the BEC based model system. In this section, we shall consider the case where the BECs are flowing which are known as atom lasers and we also look for a more novel scenario by investigating if two parallel beams of moving phonons actually deflect from its path due to gravitational effects or graviton exchange between the two beams. In case of comoving particles the geodesic deviation equation is given by
\begin{equation}\label{4.30}
\frac{d^2 x^\mu}{d\tau^2}+\Gamma^{\mu}_{~\alpha\beta}\frac{dx^\alpha}{d\tau}\frac{dx^\beta}{d\tau}=0~.
\end{equation}
For a parallel pulse, the gravitational fluctuation in the harmonic gauge reads
\begin{equation}\label{4.31}
h_{\mu\nu}=\frac{4G}{c^4}\int \frac{T_{\mu\nu}(t-|\vec{r}-\vec{r}'|,r')}{|\vec{r}-\vec{r}'|}d^3\vec{r}'
\end{equation}
For an electromagnetic pulse moving along the $z$ direction $k^{\mu}=\{1,0,0,1\}$ and ($k_{\mu}=\{-1,0,0,1\}$).  Now, if we consider two comoving light rays, the propagation vectors for the waves satisfy $k_\mu k^\mu=0$ with the energy momentum tensor being $T_{\mu\nu}=\rho c^2 k_\mu k_\nu$. As a result, we obtain $h_{00}=-h_{03}=-h_{30}=h_{33}=\Phi$ (say) because $T_{00}=-T_{03}=-T_{30}=T_{33}=\rho c^2$. The deflection in the $x$ direction then simply reads
\begin{equation}\label{1.15}
\begin{split}
 \frac{d^2 x}{dt^2}
=&\frac{c^2}{2}\left(\partial_x\Phi-\partial_x\Phi-\partial_x\Phi+\partial_x\Phi\right)=0
\end{split}
\end{equation}
where $v_z=1+\mathcal{O}(h)$.
In the harmonic gauge $\Gamma$ will become zero and as a result the overall deflection is zero. In case of a Bose-Einstein condensate the phonons can be both static as well as dynamic which allows for the consideration of a different scenario. Consider the two harmonic traps from which the bosons are outcoupled to create atom lasers. For two parallel atom laser beams travelling along the $z$ direction, the wave vectors actually take the form $k_{\mu}=\{-1,0,0,\frac{v_z}{c}\}$ with $v_z\ll c$.  Here, $T_{\mu\nu}^{\text{phonons}}\simeq T_{00}\delta^0_{~\mu}\delta^0_{~\nu}$ for the static case but for atom lasers freely falling under the effect of Earth' s gravitational field, we need to consider other components as well. Hence, the deviation along the direction $x^i$ becomes
\begin{equation}\label{1.16}
\begin{split}
\frac{d^2 x^i}{dt^2}=&\frac{c^2}{2}(\partial_xh_{00}+2\partial_xh_{0z}+\partial_xh_{zz})\\
=& 8\pi N_0 m G x^i \left(\frac{m\omega}{\pi\hbar}\right)^{\frac{3}{2}}e^{-\frac{m\omega r^2}{\hbar}}\left(1-\frac{2v_z}{c}+\frac{v_z^2}{c^2}\right)
\end{split}
\end{equation}
where $i=\{1,2\}$ showing a finite deflection of the two phonon beams travelling parallel to each other. It is evident that for two vastly separated BECs the deflection will be negligible because of the exponential suppression. The direct quantum gravitational analogue to this model will provide a standard deviation in the $x$ direction (considering $\hat{h}_{\mu\nu}=h_{\mu\nu}^{(0)}+\delta\hat{h}_{\mu\nu}$) of the order of $(\Delta x)^2\sim \langle \delta\hat{h}^2\rangle$ signifying graviton exchange between the two beams.
\section{Conclusion}\label{S5}
\noindent In this work, we consider the gravity mediated entanglement generation of phonons and compare it to the well-known quantum gravitational entanglement of masses or QGEM protocol. We start by considering two systems where the phonons get entangled by means of simple operator valued Newtonian potential  and observe the relative phase shift and the total time of entanglement generation for the model. We find out that the entanglement time is astronomically large unless the two systems has high phonon frequencies or large number of phonons with a very low separation distance. For Bose-Einstein condensates, the occurrence of large number of single mode phonon states is a natural behaviour and so we start by considering two Bose-Einstein condensates placed at two different harmonic traps separated by a finite distance $d$. We start with the basic consideration of the Gross-Pitaevski equation and considering small-amplitude oscillations, we then obtain the analytical expression for the order-parameter of the Bose-Einstein condensate where the fluctuation part is expressed in terms of the quasi-particle creation and annihilation operators. Using this analytical expression for the order parameter, we then write down the energy-momentum tensor for the model system. After several analytical steps, we finally arrive at the graviton induced shift in the energy of the gravitational field in eq.(\ref{2.24}) which is one of the main results in our paper. From there, we obtain the time required for generation of maximal entanglement and found it to be of the order of $10^{4}$ sec ($5-6$ hours time) for each condensate systems having a total of $N_0\sim10^9$ number of particles. Finally, we obtain the analytical expression for concurrence which measures the degree of entanglement between the two Bose-Einstein condensate where the entanglement is generated purely due to the interaction of the gravitons with the two identical Bose-Einstein condensates. We observe that the entanglement is generated between the phonon modes of the condensates in compared to the entanglement of masses as has been observed in \cite{Mechanism_Gravitons}. We then plot the concurrence of our model against $d$ which denotes the distance between the two centre of masses of the harmonic traps inside of which the BECs are created. We compare this concurrence plot with result obtained in \cite{Mechanism_Gravitons} and we observe that if the two harmonic traps are placed very close to each other ($d\sim 10-20~\mu \text{m}$ for $\omega=10$ Hz), then even for a condensate with very small number of particles ($N_0\sim 20$), entanglement is highly amplified compared to the case of \cite{Mechanism_Gravitons}. However, with the increase in $d$, the entanglement rapidly falls of because of the exponential damping factor in our case. We also find out that in the $\sigma\rightarrow\infty$ limit or completely delocalized wave function limit our result goes to the $\mathcal{C}\sim\frac{Gm}{d^3\omega^2}$ limits. It is better to create condensates where $\omega$ is not very large as if $\omega$ is large the $d\sim \sqrt{\frac{2\hbar}{m\omega}}$ becomes very small increasing the difficulty of experimental implementation. For a true experimental implementation, one needs to create two BECs in two separate harmonic traps such that the distance $d$ can be tuned. Then properly implementing the experimental steps, one can check for graviton induced entanglement between two Bose-Einstein condensates. We have then explored the physical scenario where two parallel flowing atom laser beams get deflected due to the background fluctuations generated via the back-reaction of the energy-momentum tensor and observed a finite displacement in the direction perpendicular to the propagation vector. It is important to remember, however, that only by measuring the deflection it is not possible to determine whether the deflection is caused by the stochastic behaviour of the classical gravitational wave or graviton induced quantum noise fluctuations. We therefore leave it for a future endeavour.
 
\appendix
\onecolumngrid
\section*{Supplementary Material}
\begin{adjustwidth}{50pt}{50pt}
Here we provide detailed derivation and added explanations of some of the analytical results presented in the primary manuscript ``Gravity mediated entanglement of phonons in Bose-Einstein condensates".
\end{adjustwidth}
\section{Small amplitude oscillations and the Bogoliubov transformation}
\noindent The single particle Gross-Pitaevski equation is given by \cite{BEC_Pitaevski_Stingari}
\begin{equation}\label{2.S.1}
\left(-\frac{\hbar^2}{2m}\frac{\partial^2}{\partial \vec{r}^2}+\frac{1}{2}m \omega^2 \vec{r}^2+\frac{4\pi\hbar^2 \mathcal{a}}{m}|\psi(\vec{r})|^2\right)\psi(\vec{r})=\mu \psi(\vec{r})~.
\end{equation}
For one spatial direction, the wave function can simply be expressed as
\begin{equation}\label{2.S.2}
\psi_0(x)=\left(\frac{m\omega}{\pi \hbar}\right)^{\frac{1}{4}}\exp\left[-\frac{m\omega x^2}{2\hbar}\right]
\end{equation}
where $\mu=\frac{1}{2}\hbar\omega$ and $g$ is set to zero.
  A complete analysis using a perturbation theory gives the analytical form of the ground state wave function up to $\mathcal{O}(g)$ to be 
\begin{equation}\label{2.S.3}
\psi^g_0(x)=\psi^{(0)}_0(x)+g\sum_{n\neq0}C_n\psi_n^{(0)}(x)
\end{equation}
where the analytical form of $C_n$ is given by
\begin{equation}\label{2.S.4}
\begin{split}
C_n&=\frac{1}{g\left(E_0^{(0)}-E_n^{(0)}\right)}\int_{-\infty}^{\infty}dx~ \psi_n^{(0)}(x)g\psi_0^{(0)^2}(x)\psi_0^{(0)}(x)\\
&=-\frac{1}{\pi n\hbar \omega}\sqrt{\frac{m\omega}{2^nn!\hbar}}\int_{-\infty}^{\infty}d\kappa ~H_n(\kappa)~e^{-2\kappa^2}
\end{split}
\end{equation}
where $\kappa=\sqrt{\frac{m\omega}{\hbar}}x$. For a near ideal Bose gas the effect of $g$ can indeed be neglected by approximation the condensate wave-function with $\psi_0^{(0)}(x)$ only\footnote{Here the analytical expressions for the unperturbed energy of the ground state and $n$-th excited state for the system are given by $E_0^{(0)}=\frac{\hbar \omega}{2}$ and $E_n^{(0)}=\left(n+\frac{1}{2}\right)\hbar\omega$}. It is important that the above wave function is more suitable for Bose-Einstein condensates of dilute gasses where the two-particle interaction is almost negligible.
For a bosonic system, as we have already discussed in the main manuscript, the field operator $\hat{\psi}(r)$ can be expressed as
\begin{equation}\label{2.S.5}
\hat{\psi}(\vec{r})=\sum_k\hat{a}_k\psi_k=\hat{a}_0\psi_0+\sum_{k\neq 0}\hat{a}_k\psi_k
\end{equation}
For a BEC one can simply use the substitution $\hat{a}_0\sim\sqrt{N_0}$ which recasts the order parameter as
\begin{equation}\label{2.S.8}
\hat{\psi}(\vec{r})=\sqrt{N}_0\psi_0(\vec{r})+\delta \hat{\psi}(\vec{r})~.
\end{equation}
For small amplitude oscillations, it is possible to express the time-dependent order parameter in the following way
\begin{equation}\label{2.S.9}
\begin{split}
\psi(\vec{r},t)&=\Psi(\vec{r},t)e^{-\frac{i\mu t}{\hbar}}\\&=\left(\Psi_0(\vec{r})+\delta\Psi(\vec{r},t)\right)e^{-\frac{i\mu t}{\hbar}}
\end{split}
\end{equation}
with $\mu$ being the chemical potential. It is important that the time dependent fluctuation terms are treated classically which shall later be raised to operator status. The small term $\delta\Psi(\vec{r},t)$ can be expanded as \cite{BEC_Pitaevski_Stingari}
\begin{equation}\label{2.S.10}
\delta\Psi(\vec{r},t)=\sum_k\left(u_k(\vec{r})e^{-i\omega_k t}+v_k^*(\vec{r})e^{i\omega_k t}\right)
\end{equation}
where $\omega_k$ is the frequency with $u_k(\vec{r})$ and $v_k^*(\vec{r})$ being the Bogoliubov coefficients corresponding to the $k$-th excitation. For a harmonic trap potential, we shall now try to find out the analytical expressions for the Bogoliubov coefficients in an one dimensional model. The time-dependent Schr\"{o}dinger equation in one spatial dimension reads $i\hbar\frac{\partial \psi(x,t)}{\partial t}=\hat{H}_0\psi(x,t)$ where $\hat{H}_0$ is given by $\hat{H}_0=-\frac{\hbar^2}{2m}\frac{\partial^2}{\partial x^2}+\frac{1}{2}m\omega^2x^2+g\left|\psi(x)\right|^2$. Substituting the analytical expression of $\psi(\vec{r},t)$ from eq.(\ref{2.S.9}) when $\vec{r}=\{x,0,0\}^T$, we obtain three independent equations given as
\begin{align}
\hat{H}_0\Psi_0(x)&=\mu\Psi_0(x)\label{2.S.11}\\
\hbar\omega_ku_k(x)&=(\hat{H}_0-\mu)u_k(x)\label{2.S.12}\\
-\hbar\omega_k v_k^*(x)&=(\hat{H}_0-\mu)v_k^*(x)\label{2.S.13}~.
\end{align}
In the above equations, eq.(\ref{2.S.11}) gives the Gross-Pitaevski equation in one dimensions whereas eq.(s)(\ref{2.S.12},\ref{2.S.13}) give the Bogoliubov equations. Our first aim is to obtain analytical forms of the Bogoliubov functions $u_k(x)$ and $v_k^*(x)$ for a Bose-Einstein condensate placed inside of a harmonic trap potential. Taking the very weak interaction limit, that is $g\rightarrow 0$, we can recast eq.(s)(\ref{2.S.12},\ref{2.S.13}) as
 \begin{align}
 \frac{\partial^2u_k(\xi)}{\partial\xi^2}-\xi^2u_k(\xi)+\left(2n_k+1\right)u_k(\xi)=&0\label{2.S.14}\\
  \frac{\partial^2v_k^*(\xi)}{\partial\xi^2}-\xi^2v_k^*(\xi)+\left(1-2n_k\right)v_k^*(\xi)=&0\label{2.S.15}
 \end{align}
 where $\xi\equiv\sqrt{\frac{m\omega}{\hbar}}x$ and $n_k\equiv \frac{\omega_k}{\omega}$. For very low energy fluctuations, we obtain $n_k\rightarrow 0$ limit which gives us the standard analytical forms for both $u_k(x)$ and $v_k^*(x)$ to be equal to $\psi_0(x)$ from eq.(\ref{2.S.4}) up to some initial coefficient. For convenience, we choose $\omega_k$ such that $n_k\in \mathbb{Z}$. To solve eq.(\ref{2.S.14}), we make a change of variables as $u_k(\xi)=\zeta_u(\xi)e^{-\frac{\xi^2}{2}}$. Substituting this ansatz back in eq.(\ref{2.S.14}), we obtain the following equation
 \begin{equation}\label{2.S.16}
 \frac{\partial^2\zeta_u(\xi)}{\partial\xi^2}-2\xi \frac{\partial\zeta_u(\xi)}{\partial\xi}+2n_k\zeta_\mu(\xi)=0
 \end{equation}
 and a generic solution of the above equation reads
 \begin{equation}\label{2.S.17}
 \zeta_\mu(\xi)=H_{n_k}(\xi)
 \end{equation}
 and the complete solution is then given by 
 \begin{equation}\label{2.S.18}
 u_k(\xi)=H_{n_k}(\xi)e^{-\frac{\xi^2}{2}}~.
 \end{equation}
 Similarly for eq.(\ref{2.S.15}) involving $v_{k}(\xi)$, we take the ansatz $v_k^*(\xi)=\zeta_v(\xi)e^{-\frac{\xi^2}{2}}$ and substitute back in eq.(\ref{2.15}). Following similar procedure, we can obtain the solution for $\zeta_v(\xi)$ and the full solution of $v_k^*(\xi)$ as
 \begin{equation}\label{2.S.19}
 \begin{split}
 v_k^*(\xi)=H_{-n_k}(\xi)e^{-\frac{\xi^2}{2}}.
 \end{split} 
 \end{equation}
Substituting eq.(s)(\ref{2.S.18},\ref{2.S.19}) back in eq.(\ref{2.S.10}), we can express the fluctuation term for all possible mode frequencies as (considering only the $x$ direction)
\begin{equation}\label{2.S.20}
\delta\Psi(x,t)=\sum\limits_k\left(H_{n_k}\left(\sqrt{\frac{m\omega}{\hbar}}x\right)e^{-\frac{m\omega x^2}{2\hbar}}e^{-i\omega_k t}+H_{-n_k}\left(\sqrt{\frac{m\omega}{\hbar}}x\right)e^{\frac{m\omega x^2}{2\hbar}}e^{i\omega_k t}\right)~.
\end{equation}
Raising the fluctuation term to operator status and combining it with the condensate part of the solution, we can write down the analytical expression for the order parameter as
\begin{equation}\label{2.S.21}
\begin{split}
\hat{\Psi}(x,t)=&\left(\Psi_0(x)+\delta\hat{\Psi}(x,t)\right)e^{-\frac{i\mu t}{\hbar}}\\
=&\left(\sqrt{N_0}\psi_0(x)+\zeta\sum\limits_k\left(H_{n_k}\left(\sqrt{\frac{m\omega}{\hbar}}x\right)e^{-\frac{m\omega x^2}{2\hbar}}e^{-i\omega_k t}\hat{b}_k+H_{-n_k}\left(\sqrt{\frac{m\omega}{\hbar}}x\right)e^{\frac{m\omega x^2}{2\hbar}}e^{i\omega_k t}\hat{b}^\dagger_k\right)\right)e^{-\frac{i\mu t}{\hbar}}
\end{split}
\end{equation}
with $\zeta$ being a constant which is proposition
We now restrict ourselves to a single mode only and
as we are primarily considering small fluctuations, we can choose $\omega_k$ to be very small such that $n_k=\frac{\omega_k}{\omega}\rightarrow 0$. This also implies that $\hbar\omega_k\ll \mu$. As discussed earlier, this limit gives the very low energy fluctuation of the overall condensate model. This collective small amplitude fluctuations also describe Bogoliubov-quasiparticles or phonons. For a fixed $k$, we can express the quasiparticle vacuum $|0\rangle_k$ as
\begin{equation}\label{2.S.22}
\hat{b}_k|0\rangle_k=0~.
\end{equation}
The raising and lowering operators satisfy the following commutation realtions
\begin{equation}\label{2.S.23}
\begin{split}
[\hat{b}_k,\hat{b}^\dagger_{k'}]=\delta_{k,k'}~,~~[\hat{b}_k,\hat{b}_{k'}]=[\hat{b}_k^\dagger,\hat{b}_{k'}^\dagger]=0~.
\end{split}
\end{equation}
As $n_k\rightarrow 0$, $H_{\pm n_k}\left(\sqrt{\frac{m\omega}{\hbar}}x\right)\rightarrow 1$. As $\hbar\omega_k\ll\mu$, we can always simplify the expression further by neglecting $\hbar\omega_k$ with respect to $\mu$. Finally, considering positive energy modes only, we can finally reduce the expression for the order parameter $\hat{\Psi}(x,t)$ as
\begin{equation}\label{2.SS.24}
\hat{\Psi}(x,t)\simeq\left(\sqrt{N_0}\psi_0(x)+\zeta e^{-\frac{m\omega x^2}{2\hbar}}\hat{b}_k\right)e^{-\frac{i\mu t}{\hbar}}.
\end{equation}
Now, $\zeta$ can be chosen as $\zeta=\left(\frac{m\omega}{\pi\hbar}\right)^\frac{1}{4}$, which helps us to write down the position dependent part of the order parameter to be
\begin{equation}\label{2.SS.25}
\hat{\Psi}(x)=\sqrt{N_0}\left(\frac{m\omega}{\pi\hbar}\right)^\frac{1}{4}e^{-\frac{m\omega x^2}{2\hbar}}\left(1+\varepsilon\hat{b}_k\right)
\end{equation}
with the definition $\varepsilon\equiv\frac{1}{\sqrt{N_0}}$ which indeed is a very small quantity as $N_0\gg 1$. Eq.(\ref{2.SS.25}) is eq.(10) in the primary manuscript.
\section{Deriving the lowest order matter-matter interaction Hamiltonian}
\noindent The energy-momentum tensor for the system reads
\begin{equation}\label{2.S.24}
\begin{split}
\hat{T}_{00}(\vec{r})= mc^2 \hat{\Psi}^{\dagger}(\vec{r})\hat{\Psi}(\vec{r})L^3\left(\delta(\vec{r}-\hat{\vec{r}}_A)+\delta(\vec{r}-\hat{\vec{r}}_B)\right)
\end{split}
\end{equation}
with $L$ being a length scale which fixes the dimension of the energy momentum tensor. Here, $\hat{\vec{r}}_A=\{\hat{x}_A,0,0\}$ and $\hat{\vec{r}}_B=\{\hat{x}_B,0,0\}$ with the operators $\hat{x}_A$ and $\hat{x}_B$ being
$\hat{x}_A=\frac{d}{2}+\delta \hat{x}_A~\text{and}~\hat{x}_B=-\frac{d}{2}+\delta\hat{x}_B.$
The fluctuation operators $\delta \hat{x}_A$ and $\delta\hat{x}_B$ are representable in terms of the raising and lowering operators of the phonons at the positions $A$ and $B$ respectively. The fluctuations $\delta\hat{x}_A$ and $\delta\hat{x}_B$ in terms of the phonon-raising and lowering operators read
$\delta\hat{x}_A=\sqrt{\frac{\hbar}{2m\omega}}(\hat{\alpha}+\hat{\alpha}^{\dagger})~\text{and}~\delta\hat{x}_B=\sqrt{\frac{\hbar}{2m\omega}}(\hat{\beta}+\hat{\beta}^{\dagger})~.$
 Using eq.(\ref{2.S.24}) and the property of the Dirac-delta function ($f(x)(x-a)=f(a)$) along with the expressions for the quantum fluctuations, we can write down the expression of the energy-momentum tensor as
\begin{equation}\label{2.47}
\begin{split}
\hat{T}_{00}(\vec{r})=N_0mc^2\left(\frac{m\omega}{\pi\hbar}\right)^{\frac{3}{2}}L^3\left[(1+\varepsilon \hat{\alpha}^\dagger)e^{-\frac{m\omega \hat{r}_A^2}{\hbar}}(1+\varepsilon\hat{\alpha})\delta(\vec{r}-\hat{\vec{r}}_A)\otimes\mathbb{1}_B+\mathbb{1}_A\otimes(1+\varepsilon \hat{\beta}^\dagger)e^{-\frac{m\omega \hat{r}_A^2}{\hbar}}(1+\varepsilon\hat{\beta})\delta(\vec{r}-\hat{\vec{r}}_B)\right]~.
\end{split}
\end{equation} 
The Fourier transform of the energy-momentum tensor takes the analytical form
\begin{equation}\label{2.48}
\hat{T}_{00}(\vec{k})=\frac{1}{2\pi}\int d^3\vec{r}~\hat{T}_{00}(\vec{r})e^{-i\vec{k}\cdot\vec{r}}~.
\end{equation}
After being acted on by the Dirac delta functions, the exponential term will get operator valued and as a result we shall need to implement Weyl ordering to get rid of any inconsistencies. The Weyl ordered Fourier transformation of the energy momentum tensor reads
\begin{equation}\label{2.49}
\hat{T}_{00}(\vec{k})=\frac{\mathcal{A}_{N_0}}{(2\pi)^{\frac{3}{2}}}\left[e^{-\frac{i\vec{k}\cdot\hat{\vec{r}}_A}{2}}(1+\varepsilon\hat{\alpha}^\dagger)e^{-\frac{m\omega \hat{r}_A^2}{\hbar}}(1+\varepsilon\hat{\alpha})e^{-\frac{i\vec{k}\cdot\hat{\vec{r}}_A}{2}}\otimes \hat{\mathbb{1}}_B+\hat{\mathbb{1}}_A\otimes e^{-\frac{i\vec{k}\cdot\hat{\vec{r}}_B}{2}}(1+\varepsilon\hat{\beta}^\dagger)e^{-\frac{m\omega \hat{r}_B^2}{\hbar}}(1+\varepsilon\hat{\beta})e^{-\frac{i\vec{k}\cdot\hat{\vec{r}}_B}{2}}\right]
\end{equation}
with $\mathcal{A}_{N_0}$ being defined as $\mathcal{A}_{N_0}\equiv N_0mc^2L^3\left(\frac{m\omega}{\pi\hbar}\right)^{\frac{3}{2}}=N_0mc^2\left(\frac{m\omega L^2}{\pi\hbar}\right)^{\frac{3}{2}}$.
Here, our aim is to investigate gravity mediate entanglement between the condensates at positions $A$ and $B$, and we shall restrict ourselves to the $|1_A,1_B\rangle$ states only where any higher order states like $|n_A,n_B\rangle$ where $n_A>1_A$ or $n_B>1_B$ is not considered. The terms that can contribute towards entanglement generation between the two-condensates at positions $A$ and $B$ can be separated and the contributing terms of the interaction Hamiltonian can be expressed as
\begin{equation}\label{2.50}
\begin{split}
\Delta\hat{H}_g^{AB}=-\frac{\mathcal{A}_{N_0}^2\kappa^2}{8c^2(2\pi)^2}\int_0^\infty dk\int_0^\pi d\theta \sin\theta ~\biggr[e^{\frac{i\vec{k}\cdot\hat{\vec{r}}_A}{2}}(1+\varepsilon\hat{\alpha}^\dagger)e^{-\frac{m\omega \hat{r}_A^2}{\hbar}}(1+\varepsilon\hat{\alpha})e^{\frac{i\vec{k}\cdot(\hat{\vec{r}}_A-\hat{\vec{r}}_B)}{2}}(1+\varepsilon\hat{\beta}^\dagger)e^{-\frac{m\omega \hat{r}_B^2}{\hbar}}(1+\varepsilon\hat{\beta})e^{-\frac{i\vec{k}\cdot\hat{\vec{r}}_B}{2}}&\\+e^{\frac{i\vec{k}\cdot\hat{\vec{r}}_B}{2}}(1+\varepsilon\hat{\beta}^\dagger)e^{-\frac{m\omega \hat{r}_B^2}{\hbar}}(1+\varepsilon\hat{\beta})e^{\frac{i\vec{k}\cdot(\hat{\vec{r}}_B-\hat{\vec{r}}_A)}{2}}(1+\varepsilon\hat{\alpha}^\dagger)e^{-\frac{m\omega \hat{r}_A^2}{\hbar}}(1+\varepsilon\hat{\alpha})e^{-\frac{i\vec{k}\cdot\hat{\vec{r}}_A}{2}}\biggr]&
\end{split}
\end{equation}
where the $\phi$ integral has been executed. The next step is to execute the theta integrals. As only the $x$ components of both the $\hat{\vec{r}}_A$ and $\hat{\vec{r}}_B$ operators are non-vanishing, we consider that they will create the same angle with the wave vector $\vec{k}$. Hence, we can write safely $\exp[i\vec{k}\cdot\hat{\vec{r}}_A]=\exp[ik|\hat{\vec{r}}_A|\cos\theta]$ and $\exp[i\vec{k}\cdot\hat{\vec{r}}_B]=\exp[ik|\hat{\vec{r}}_B|\cos\theta]$. In order to execute the $\theta$ integral, we need to shift the phase terms to the right. We can obtain following simple results after doing a little bit of analytical simplifications
\begin{align}
\exp\left[\frac{i\vec{k}\cdot\hat{\vec{r}}_A}{2}\right]\hat{\alpha}^\dagger&=\left(\hat{\alpha}^\dagger-ik\rho\cos\theta\right)\exp\left[\frac{i\vec{k}\cdot\hat{\vec{r}}_A}{2}\right]\label{2.51}\\
\exp\left[\frac{i\vec{k}\cdot\hat{\vec{r}}_A}{2}\right]\hat{\alpha}&=\left(\hat{\alpha}+ik\rho\cos\theta\right)\exp\left[\frac{i\vec{k}\cdot\hat{\vec{r}}_A}{2}\right]\label{2.52}
\end{align}
\linebreak
where $\rho\equiv \frac{1}{2}\sqrt{\frac{\hbar}{2m\omega}}$. Similarly, for the counterparts of eq.(s)(\ref{2.51},\ref{2.52}) using the $\beta$ and $\beta^\dagger$ operators read
\smallskip
\begin{align}
\exp\left[\frac{i\vec{k}\cdot\hat{\vec{r}}_B}{2}\right]\hat{\beta}^\dagger&=\left(\hat{\beta}^\dagger+ik\rho\cos\theta\right)\exp\left[\frac{i\vec{k}\cdot\hat{\vec{r}}_B}{2}\right]\label{2.53}
\end{align}
and
\begin{align}
\exp\left[\frac{i\vec{k}\cdot\hat{\vec{r}}_B}{2}\right]\hat{\beta}&=\left(\hat{\beta}-ik\rho\cos\theta\right)\exp\left[\frac{i\vec{k}\cdot\hat{\vec{r}}_B}{2}\right]\label{2.54}~.
\end{align} 

\noindent We shall now make use of eq.(s)(\ref{2.51},\ref{2.52},\ref{2.53},\ref{2.54}) to simplify and shift the phase terms to right in the expression of the interaction part of the Hamiltonian $\Delta\hat{H}_g^{AB}$ in eq.(\ref{2.50}).
Up to $\mathcal{O}(\varepsilon^2)$ one can then express eq.(\ref{2.50}) as
\begin{equation}\label{2.55}
\begin{split}
\Delta\hat{H}_g^{AB}=&-\frac{\mathcal{A}_{N_0}^2\kappa^2}{8c^2(2\pi)^2}\int_0^\infty dk\int_0^\pi d\theta \sin\theta\biggr[\left(\left(\left(1+2k^2\rho^2\xi^2\cos^2\theta\right)
+\varepsilon\left(\hat{\alpha}^\dagger+\beta^\dagger\right)++\varepsilon^2\hat{\alpha}^\dagger\hat{\beta}^\dagger\right)e^{-\frac{m\omega}{\hbar}\left(\hat{r}_A^2+\hat{r}_B^2\right)}+\cdots\right)\\\times&\cos\left[k|\hat{\vec{r}}_B-\hat{\vec{r}}_A|\cos\theta\right]+k\rho\varepsilon^2\cos\theta\left(\left(\hat{\alpha}^\dagger+\beta^\dagger\right)e^{-\frac{m\omega}{\hbar}\left(\hat{r}_A^2+\hat{r}_B^2\right)}-e^{-\frac{m\omega}{\hbar}\left(\hat{r}_A^2+\hat{r}_B^2\right)}(\hat{\alpha}+\hat{\beta})\right)\sin\left[k|\hat{\vec{r}}_B-\hat{\vec{r}}_A|\cos\theta\right]\biggr].
\end{split}
\end{equation}
We are now in a position to execute the $\theta$ and consecutively the $k$ integrals. To obtain the analytical expression of the interaction Hamiltonian, we need to execute the following three integrals given as
\begin{align}
\mathcal{I}_1=&\int_0^\infty dk\int_0^\pi d\theta \sin\theta \cos(k|\hat{\vec{r}}_B-\hat{\vec{r}}_A|\cos\theta)\label{2.56}\\
\mathcal{I}_2=&\int_0^\infty dk k\int_0^\pi d\theta \sin\theta \cos\theta\sin(k|\hat{\vec{r}}_B-\hat{\vec{r}}_A|\cos\theta)\label{2.57}\\
\mathcal{I}_3=&\int_0^\infty dk k^2\int_0^\pi d\theta \sin\theta\cos^2\theta \cos(k|\hat{\vec{r}}_B-\hat{\vec{r}}_A|\cos\theta)~.\label{2.58}
\end{align}
One can simply obtain the analytical form of the integral in eq.(\ref{2.56}) as ($\lambda=\cos\theta$)
\begin{equation}\label{2.59}
\begin{split}
\mathcal{I}_1=&\int_0^\infty dk\int_{-1}^{1} d\lambda \cos(k|\hat{\vec{r}}_B-\hat{\vec{r}}_A|\lambda)=2\int_0^\infty dk \frac{\sin(k|\hat{\vec{r}}_B-\hat{\vec{r}}_A|\lambda)}{k|\hat{\vec{r}}_B-\hat{\vec{r}}_A|}=\frac{\pi}{|\hat{\vec{r}}_B-\hat{\vec{r}}_A|}~.
\end{split}
\end{equation}
Dropping any oscillatory contributions and using a standard regularization procedure, one can simply obtain the analytical expressions for the integrals in eq.(s)(\ref{2.57},\ref{2.58}) as
\begin{align}\label{2.60}
\mathcal{I}_2=&\frac{\pi}{|\hat{\vec{r}}_B-\hat{\vec{r}}_A|^2}~\text{and}~
\mathcal{I}_3=-\frac{2\pi}{|\hat{\vec{r}}_B-\hat{\vec{r}}_A|^3}~.
\end{align}
We shall keep terms like $\hat{\alpha}^\dagger\hat{\beta}^\dagger$ in the analytical expression for the interaction Hamiltonian. We shall now perturbatively expand the $\frac{1}{|\hat{\vec{r}}_B-\hat{\vec{r}}_A|}$ term
\begin{equation}\label{2.61}
\begin{split}
\frac{1}{|\hat{\vec{r}}_B-\hat{\vec{r}}_A|}&=\frac{1}{d+(\delta\hat{x}_B-\delta\hat{x}_A)}=\frac{1}{d}\left(1-\frac{\hbar}{m\omega d^2}(\hat{\alpha}^\dagger\hat{\beta}^\dagger+\cdots).-\sqrt{\frac{\hbar}{2m\omega d^2}}\left(\hat{\beta}^\dagger-\hat{\alpha}^\dagger+\cdots\right)+\cdots\right)~.
\end{split}
\end{equation}
Similarly, we can expand the $\frac{1}{|\hat{\vec{r}}_B-\hat{\vec{r}}_A|^2}$ and $\frac{1}{|\hat{\vec{r}}_B-\hat{\vec{r}}_A|^3}$ terms as
\begin{align}
 \frac{1}{|\hat{\vec{r}}_B-\hat{\vec{r}}_A|^2}&=\frac{1}{d^2}\left[1-\frac{3\hbar}{m\omega d^2}(\hat{\alpha}^\dagger\hat{\beta}^\dagger+\cdots)+\cdots\right]\label{2.62}\\
  \frac{1}{|\hat{\vec{r}}_B-\hat{\vec{r}}_A|^3}&=\frac{1}{d^3}\left[1-\frac{6\hbar}{m\omega d^2}(\hat{\alpha}^\dagger\hat{\beta}^\dagger+\cdots)+\cdots\right]\label{2.63}~.
\end{align}

\noindent The next important step is to understand the decomposition of the exponential term $e^{-\frac{m\omega}{\hbar}\left(\hat{r}_A^2+\hat{r}_B^2\right)}$ as follows
\begin{equation}\label{2.64}
\begin{split}
e^{-\frac{m\omega}{\hbar}(\hat{r}_A^2+\hat{r}_B^2)}&=e^{-\frac{m\omega d^2}{2\hbar}}\left(1-\frac{m\omega d^2}{\hbar}\left(\delta\hat{x}_B-\delta\hat{x}_A\right)-\frac{m\omega}{\hbar}(\delta\hat{x}_B^2+\delta\hat{x}_A^2)+\frac{1}{2}\left(\frac{m\omega d}{\hbar}\right)^2\left(\delta\hat{x}_B-\delta\hat{x}_A\right)^2+\cdots\right)~.
\end{split}
\end{equation}
As the expansion coefficient $\frac{m\omega d^2}{\hbar}$ is greater than unity, we need to keep all the terms in the operator expansion and treat it nonperturbatively. 
\noindent Using eq.(s)(\ref{2.56},\ref{2.57},\ref{2.58},\ref{2.59},\ref{2.60},\ref{2.64}) and substituting back in eq.(\ref{2.55}), we obtain the analytical form of the interaction Hamiltonian as
\begin{equation}\label{2.65}
\begin{split}
:\Delta\hat{H}_g^{AB}:&=-\frac{\mathcal{A}_{N_0}^2\kappa^2}{16\pi c^2}\frac{e^{-\frac{m\omega d^2}{2\hbar}}}{d}\left[\left[1+\left[\varepsilon^2-\frac{m\omega d^2}{2\hbar}-\frac{\hbar}{m\omega d^2}\right]\hat{\alpha}^\dagger\hat{\beta}^\dagger+\cdots\right]+\frac{4\rho^2\varepsilon^2}{d^2}\left[1-\left[\frac{m\omega d^2}{2\hbar}+\frac{6\hbar}{m\omega d^2}\right]\hat{\alpha}^\dagger\hat{\beta}^\dagger\right]+\cdots\right]
\end{split}
\end{equation}
where we have implemented normal ordering in the final expression. The lowest order matter-matter interaction will be generated by the term 
\begin{equation}\label{2.66}
\begin{split}
\Delta\hat{\mathcal{H}}_g^{AB}=&-\frac{\mathcal{A}_{N_0}^2\kappa^2}{16\pi c^2}\frac{e^{-\frac{m\omega d^2}{2\hbar}}}{d}\left[\varepsilon^2-\frac{m\omega d^2}{2\hbar}-\frac{\hbar}{m\omega d^2}+\frac{4\rho^2\varepsilon^2}{d^2}\left[\frac{m\omega d^2}{2\hbar}+\frac{6\hbar}{m\omega d^2}\right]\right]\hat{\alpha}^\dagger\hat{\beta}^\dagger~.
\end{split}
\end{equation}
The coefficient $C_{1_A1_B}=\frac{\langle 1_A,1_B|:\Delta\hat{H}_g^{AB}:|0_A,0_B\rangle}{2E_0-E_A-E_B}$ can be obtained as
\begin{equation}\label{2.67}
\begin{split}
C_{1_A1_B}=&\frac{\langle 1_A,1_B|:\Delta\hat{H}_g^{AB}:|0_A,0_B\rangle}{2E_0-E_A-E_B}\\=&\frac{\langle 1_A,1_B|:\Delta\hat{\mathcal{H}}_g^{AB}:|0_A,0_B\rangle}{2E_0-E_A-E_B}+0\\
=&-\frac{\mathcal{A}_{N_0}^2\kappa^2}{32\pi \hbar\omega d c^2}e^{-\frac{m\omega d^2}{2\hbar}}\left[\frac{m\omega d^2}{2\hbar}-\frac{5\varepsilon^2}{4}+\frac{\hbar}{m\omega d^2}\left(1-\frac{3\varepsilon^2\hbar}{m\omega d^2}\right)\right]
\end{split}
\end{equation}
where we can safely neglect $\mathcal{O}\left(\frac{\hbar^2}{m^2\omega^2d^4}\right)$ terms as it will have minor contributions towards the overall entanglement generation. 
\end{document}